\documentclass[aip,rsi,reprint,graphicx]{revtex4-1} 
\usepackage{graphicx}
\usepackage[utf8]{inputenc}
\usepackage{endnotes}
\usepackage{epstopdf}

\draft 

\begin{document}

\title{Optical microscope and tapered fiber coupling apparatus for a dilution refrigerator} 

\author{A.J.R. MacDonald}
\email{ajm1@ualberta.ca}
\author{G.G. Popowich}
\author{B.D. Hauer}
\author{P.H. Kim}
\author{A. Fredrick}
\author{X. Rojas}
\author{P. Doolin}
\author{J.P. Davis}
\email{jdavis@ualberta.ca}
\affiliation{Department of Physics, University of Alberta, Edmonton, AB, Canada T6G 2E9}

\date{\today}

\begin{abstract}
 We have developed a system for tapered fiber measurements of optomechanical resonators inside a dilution refrigerator, which is compatible with both on- and off-chip devices. Our apparatus features full three-dimensional control of the taper-resonator coupling conditions enabling critical coupling, with an overall fiber transmission efficiency of up to 70\%. Notably, our design incorporates an optical microscope system consisting of a coherent bundle of 37,000 optical fibers for real-time imaging of the experiment at a resolution of $\sim$1 $\mu$m. We present cryogenic optical and optomechanical measurements of resonators coupled to tapered fibers at temperatures as low as 9 mK.
\end{abstract}


\maketitle

\section{Introduction}

Recently, the field of cavity optomechanics \cite{Aspelmeyer2013,Kippenberg2008} has attracted a great deal of attention for its promise in areas ranging from application-driven efforts to fundamental tests of quantum mechanics at the mesoscale. Nano- and micro-mechanical devices coupled to microwave and optical-frequency cavities have been shown to be exquisite force, \cite{Gavartin2012,Miao2012,Doolin2014} position, \cite{Anetsberger2010} acceleration \cite{Krause2012} and torque \cite{Kim2013,Wu2014} sensors on small scales. These properties make such systems ideal for hybrid quantum information processing and communications infrastructures where the mechanical element can be adapted to act as a transducer between photons and other physical systems such as superconducting qubits. \cite{Stannigel2010,OConnell2010,Bochmann2013} The optomechanical interaction can also be exploited to prepare exotic and non-classical states such as squeezed states, \cite{SafaviNaeini2013} entanglement \cite{Palomaki2013} and mechanical Fock states. \cite{Borkje2014} Additionally, there is significant interest in observing the quantum mechanical nature of a mesoscopic system - something that optomechanical systems are well-suited to accomplish by preparation of the mechanical element in its motional ground state.

A prerequisite to many of these quantum optomechanical experiments is to reduce the thermal phonon occupation of the system, as thermomechanical noise will drown out the signal of any quantum mechanical behaviour. With the advent of laser-cooling, it is now much more feasible to reduce the average phonon occupation to very close to the ground state. \cite{Schliesser2008,Chan2011,Teufel2011,Park2009} However, laser cooling reduces the mechanical quality factor and is limited by the temperature of the equilibrium thermal bath, \cite{Aspelmeyer2013} making cryogenic passive pre-cooling a necessity. Cryogenic optomechanics is equally interesting for its projected use in sensitively probing phenomena such as superconductivity and superfluidity, \cite{Sun13,DeLorenzo2014} which emerge only at low temperatures.

Several groups have successfully incorporated cooling systems into their optomechanical experiments, but most of these have used either helium flow cryostats \cite{Srinivasan2007,Park2009,Riviere2013} which are limited in base temperature to $>$1 K, or microwave resonators \cite{OConnell2010,Teufel2011,Palomaki2013} in which the electromagnetic coupling can be completely integrated on-chip. A photonic crystal optomechanical device has recently been thermalized to 270 mK inside a dilution fridge, \cite{Meenehan2014} using a lensed fiber to couple light to on-chip photonic crystal resonators. However, the method used limits accessible devices to a small fraction of a two-dimensional chip and has an efficiency limited to 35\%. Here, we present a system for tapered fiber coupling to both on- and off-chip optomechanical devices, achieving an overall fiber efficiency of up to 70\% in the cryostat and featuring a simple optical system to image the experiment in real-time at mK temperatures with negligible heating.

The rest of the paper is outlined as follows: Section \ref{samples} describes the optomechanical devices studied in our dilution fridge, Section \ref{CouplingApparatus} outlines the tapered fiber coupling apparatus and Section \ref{Imaging} describes the microscope design. We conclude by presenting optical and optomechanical results obtained with this system in Section \ref{results}.

\section{Optomechanical Resonators}
\label{samples}

In the optomechanical experiments described here, motion of a mechanical resonator is coupled to the field inside an optical whispering gallery mode (WGM) resonator through a position-dependent optical resonance frequency $\omega(x)=\omega_{0}-Gx$. Here $x$ is the position of the mechanical resonator, $\omega_{0}$ is the unperturbed optical resonance frequency and $G=-\partial\omega/\partial x$ is the optomechanical coupling strength. The time-varying position of a mechanical resonator can then be transduced as an oscillating amplitude or phase of the light emerging from the cavity.

We study both lithographically fabricated on-chip and laser-fabricated silica optomechanical resonators coupled to dimpled and straight tapered fibers, respectively. The dimpled fiber \cite{Hauer2014} allows us to selectively couple light into a single on-chip silicon microdisk which is itself side-coupled via its evanescent field to a mechanical element, such as a cantilever \cite{Doolin2014} or torsional paddle. \cite{Kim2013} Silica bottle resonators \cite{Kakarantzas2001,Pollinger2009} are created by simultaneous stretching and heating of an optical fiber with a CO$_{2}$ laser. The exact shape of these resonators can be varied from parabolic to spheroidal by varying the intensity of the CO$_{2}$ laser spot and the stretching time. By pulling the ends of the fiber for a long period of time, a series of bottles with very thin connecting stems are created. These bottles exhibit high-quality optical WGMs and mechanical breathing modes, which permit observation of mechanical motion at cryogenic temperatures.

\section{Coupling Apparatus}
\label{CouplingApparatus}

\begin{figure}[t]
	\includegraphics{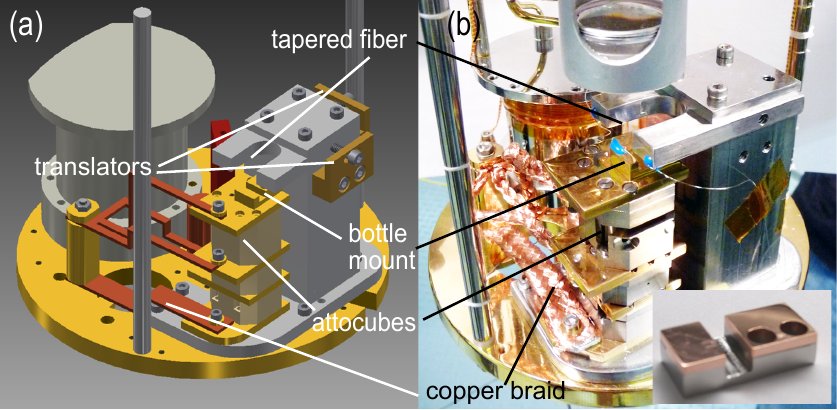}
	\caption{Coupling Apparatus. (a) Rendering and (b) photograph of the apparatus used to couple light from a straight or dimpled tapered optical fiber into an optomechanical resonator. Shown are the attocube nanopositioning stages, sample and fiber mounts, and thermal anchoring to the mixing chamber stage. The brass translators shown on the fiber mount in (a) but not pictured in (b) are for coarse alignment of the fiber with respect to the microscope as described in Section \ref{Imaging} and are removed before cooling the fridge. Inset: Copper and Invar bottle mount prior to gold-plating. \label{apparatus}}
\end{figure}

Our coupling scheme, shown in Figure \ref{apparatus}, consists of an optomechanical resonator mounted on a set of three attocube (LT-UHV) linear nanopositioning stages next to a platform for mounting a tapered optical fiber. The entire apparatus is mounted to a large Invar base, situated on the mixing chamber plate of a commercial Oxford Instruments Kelvinox 400HA dilution fridge. The use of Invar, a nickel/iron alloy which undergoes very small thermal contraction when cooled from room to cryogenic temperatures, minimizes drifts in the relative taper-sample position. We use tapered fibers to inject light into our devices for their superior efficiency and phase-matching properties  \cite{Knight1997,Cai2000,Srinivasan2005} over methods such as prism coupling \cite{Park2007,Treussart1998}, grating couplers \cite{Sun13} and lensed-fiber coupling. \cite{Meenehan2014}

We fabricate our cryogenic tapers \cite{Riviere2013,Takashima2010} by heating and stretching single mode optical fibers until they have a tapered region with a very narrow core ($\sim$1 $\mu$m). \cite{Hauer2014} This method produces efficient ($\geq$ 90\% transmission at room temperature) and adiabatic tapers in which a large fraction of the guided mode propagates in the evanescent field outside the core. Special attention is paid to the length of acrylic coating stripped from the fiber before tapering, as the acrylic protects the fiber while it is bent in the tight constraints of the fridge.

Once a suitable taper has been fabricated, it is glued to an Invar fiber-holder (Figure \ref{apparatus}) in a two-step process and then mounted to an Invar block on the Invar base plate. The fiber is first glued with a five-minute epoxy which increases the tension of the fiber as it sets. Since this epoxy will not remain adhered to the fiber and mount at low temperatures, we cover it with a low-temperature-compatible Tra-Bond epoxy which is de-gassed and allowed to set for 24 hours. 

The taper is then fusion-spliced to two lengths of fiber anchored around the still plate. The extra length of fiber is secured with Kapton and aluminum tape to the mixing chamber stage to minimize vibrations and accidental contact of loose fibers with the radiation shield of the fridge, which could cause a heat leak. Surface contaminants (primarily water) which adsorb to the tapered region and scatter light out of the guided mode can be removed by pumping on the inner vacuum can (IVC). In addition cleaning the surroundings inside the fridge dramatically improves the final taper transmission. This results in overall efficiencies as high as 70\%, with most of the loss originating from the taper and splices.

Finally the optomechanical device is mounted on the nanopositioning stages. On top of each stage is an oxygen-free high-conductivity (OFHC) copper plate which is gold-plated and thermally anchored to the mixing chamber via an OFHC copper braid (see Figure \ref{apparatus}). Bottle resonators are glued using the procedure described above to the mount shown (prior to gold-plating) in the inset of Figure \ref{apparatus}. Two pieces of OFHC copper are pushed through a long Invar piece, which is then gold-plated. The copper and gold-plating maximize thermal contact to the fridge while the Invar minimizes the effects of thermal contraction which can detension the bottles and result in low-frequency vibrations as the attocubes are moved at low temperatures.

\section{Imaging}
\label{Imaging}

\begin{figure*}[t]
	\includegraphics{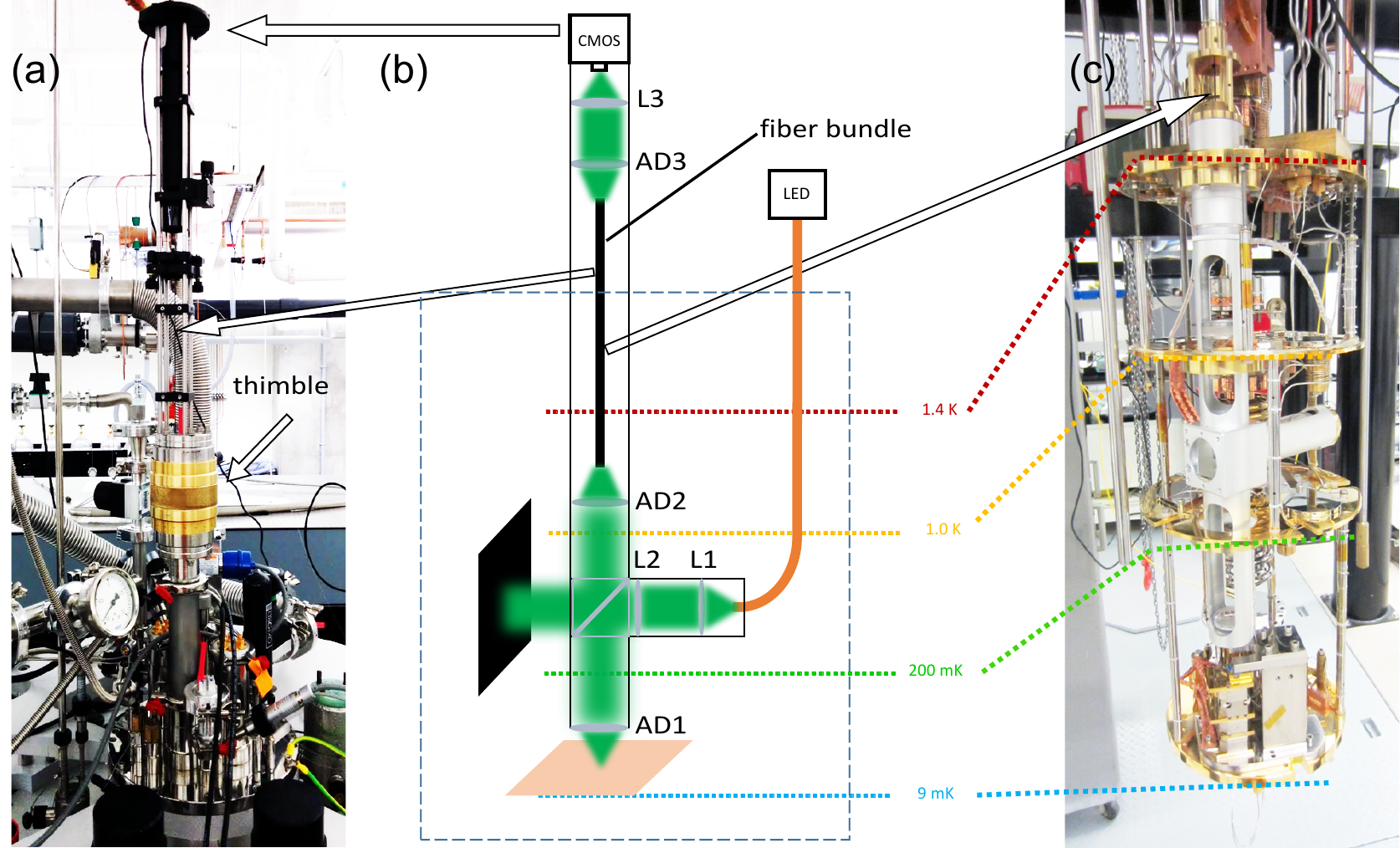}
	\caption{(a) Photograph of the imaging system above the fridge. Arrows denote the thimble used to translate the apparatus, the fiber bundle and the CMOS camera. (b) Schematic of the imaging system. The blue dashed box represents the limits of the inner vacuum can (IVC) at 4.2 K, while dotted lines represent (from top to bottom) 1 K pot, still, cold plate and mixing chamber stages. Lenses are: condenser lens (L1), long-focal length plano-convex lenses (L2, L3) and achromatic doublets (AD1, AD2, AD3). (c) Photograph of the microscope components inside the fridge. \label{endoscope}}
\end{figure*}

To efficiently couple light between the taper and the optical resonator, it is important that they are well-aligned with respect to each other. At room temperature, this can be accomplished with a relatively simple optical microscope configuration \cite{Hauer2014} but at low temperatures, an imaging system can significantly heat the environment. Differential thermal contraction of the various materials inside the fridge creates further challenges as any alignment performed at room temperature will likely shift well before the base temperature of the fridge is reached. To counteract this, we introduce a real-time imaging system which causes very little heating at low temperatures and which has enough degrees of freedom to compensate, either in real-time or before cooldown, for any shifts in the alignment.

Imaging at cryogenic temperatures has conventionally been performed in cryostats equipped with a series of windows which block room temperature radiation but allow visible light to pass. Such systems have been used successfully to image and measure $^{4}$He crystal surfaces, \cite{Balibar2005,Franck1986,Nomura2003} orientations \cite{Rojas2010} and acoustic waves. \cite{Souris2011} These optical access cryostats are however limited in their achievable base temperatures: helium flow cryostats and closed cycle fridges have base temperatures of $\sim$1 K. Dilution refrigerators are also limited by the heat load from the windows, although temperatures as low as 10 mK \cite{Endnote} can be reached with careful attention given to filtering of radio frequency (RF) and infrared radiation.

In the early 1990s, interest in directly imaging $^{3}$He and $^{4}$He at mK and sub-mK temperatures led to the development of methods to contain most of the imaging system inside the cryostat. The first such effort, at Helsinki, used a helium-neon (HeNe) laser housed at room temperature as a light source and a bundle of 30,000 optical fibers to transport an image from inside the fridge to a camera at room temperature. \cite{Manninen1992} This setup was used to interferometrically obtain the first images of superfluid $^{3}$He at temperatures down to 0.7 mK, albeit at a limited resolution. Another approach, which has become a popular technique, used a charge-coupled device (CCD) cooled to 65 K inside a dilution fridge \cite{Alles1994}, enabling imaging at temperatures below 1 mK. \cite{Wagner1994} However, this method leads to additional heating from thermal and RF radiation of the CCD as well as time delays between the room temperature control electronics and the CCD, requiring further shielding and custom timing electronics. 

We use an imaging system, similar to that of Reference \citenum{Manninen1992} and pictured in Figure \ref{endoscope}, to assist in aligning our optomechanical devices \emph{in situ}. It uses a 530 nm light-emitting diode (LED) as an illumination source, a coherent bundle of 37,000 optical fibers to transport an image to room temperature and the complimentary metal-oxide-semiconductor (CMOS) chip from a commercial webcam to construct a real-time image. Green light is brought into the fridge via a multimode fiber and collimated with a combination of strong condenser (L1 in Figure \ref{endoscope}) and weak plano-convex (L2) lenses. A cube beamsplitter reflects 10\% of the light onto the sample and dumps the remaining 90\% to a black metal velvet sheet (Acktar Advanced Coatings) affixed to the interior of the IVC, providing a heat dump. This beamsplitter configuration ensures that most of the light reflected from the sample is used for the image. Achromatic doublets focus the light onto the sample (AD1), into the bundle (AD2) and collimate the light exiting the bundle (AD3). A final plano-convex lens (L3), focuses the image onto the CMOS chip. All optics are housed in 1-inch diameter aluminum lens tubes, which are thermally anchored to the still by a copper braid. Room-temperature alignment of the illumination fiber and the fiber bundle is achieved through a pair of home-built brass xy tilt manipulators.

The portion of the microscope residing inside the fridge is suspended from above by a length of thin-walled stainless steel tubing and can be moved vertically with a 50 $\mu$m resolution over a range of 50 mm by a thimble (Huntington VF-178-275) on top of the fridge. Lens tubes are guided at each stage by a teflon ring. This mechanism allows the focal plane of the microscope to be translated without tilting or vibrating, and minimizes the heat leaks between stages. 

As cooldown proceeds, thermal contractions cause the taper and sample to drift relative to the microscope. Although we have the ability to adjust the focal plane of the system (via the thimble) as well as the sample's position (via the nanopositioning stages), it is not possible to adjust the in-plane position of the taper relative to the imaging system at low temperatures. From room to cryogenic temperatures, the taper reproducibly drifts an estimated 200 $\mu$m from the bottom right to the center of the image. Fortunately, we can compensate for this drift using the pair of home-built brass push-pull translators shown in Figure \ref{apparatus}(a) mounted on the Invar block. We position the taper a set distance outside the image at room temperature before tightening the fiber holder down and removing the translators. 

The fiber bundle and illumination fibers are brought into the fridge via a line-of-sight port and sealed with Tra-Bond epoxy at the top. To reduce thermal radiation from above, we use a series of baffles between the thimble and the stainless steel tubing, and offset the holes in each of the baffles through which the fibers pass. This ensures that there is no direct free-space line-of-sight from room temperature into the IVC. Although the fiber bundle is well-connected to room temperature, its low thermal conductivity prevents any significant conduction of heat from the outside environment. We observe a very slight increase in the base temperature corresponding to room lights ($\sim$0.1 mK as measured by a resistive thermometer), likely due to photons coupled into the fridge via the optical fibers but this is easily remedied by covering the fibers above the fridge.

\begin{figure}[t]
	\includegraphics{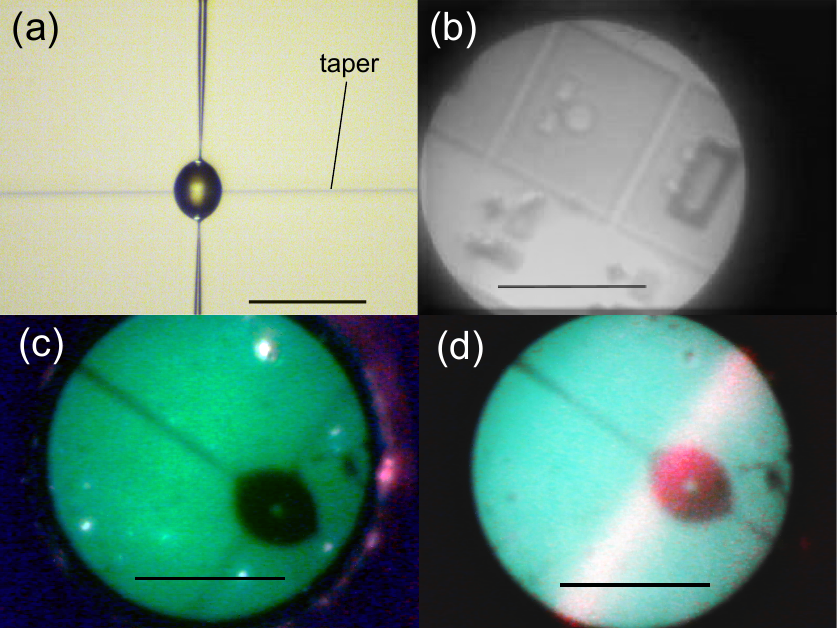}
	\caption{(a) Image of a bottle resonator and tapered fiber taken at 295 K with an optical microscope. (b) Image of an on-chip optomechanical resonator \cite{Kim2013} taken using our system at room temperature. The image has been rendered in black and white for clarity. Bottle resonator at (c) the dilution fridge base temperature of 9 mK with tapered fiber visible near the center of the bottle and (d) at 295 K with red light used to illuminate the taper. Images in (b),(c) and (d) captured with $<$1 mW of initial green LED power (see Table \ref{power}). All scale bars 100 $\mu$m. \label{Images}}
\end{figure}

Figure \ref{Images} shows images captured with our system at room and cryogenic temperatures. We achieve an effective magnification of the image by coupling only a fraction of the light reflected from the sample into the fiber bundle. Using AD1 (focal length $f=30$ mm), the green light is focused to a $\sim$1 mm spot on the sample, while AD2 ($f=100$ mm) focuses the light to a $\sim$4 mm spot at the fiber bundle - much larger than the $\sim$850 $\mu$m image circle of the bundle. Since the image is limited to just 37 kilopixels, this magnification factor of 3.3 between the sample and the fiber bundle is a compromise between a sufficient resolution to accurately position the tapered fiber and a large field of view to identify a device.

The images have a resolution of just 1 $\mu$m whereas the tapered fiber has a diameter of $\sim$1 $\mu$m, making it difficult to resolve the taper in the image. This is especially pertinent in the case of dimpled fibers where the dimple is usually found by looking for the region which comes into focus at a lower point than the rest of the fiber. \cite{Hauer2014} To facilitate this, we inject red light from a HeNe laser into the measurement fiber. The red light preferentially scatters from the dimpled or tapered region, as seen in Figure \ref{Images}(d), allowing us to more easily locate it in the image. With this adjustment we find that we can reproducibly locate the tapered fiber at low temperatures.

\begin{table}[h]
	\caption{The optical intensity of green LED light, measured in free space at the output of the LED, and the two ports of the beamsplitter. The beamsplitter reflects light onto the sample and transmits it onto the sIVC. Also shown is the measured base temperature $T$ of the fridge from which the heat load $\dot{Q}$ due to the LED light is inferred. \label{power}}
	\begin{tabular}{|c|c|c|c|c|}
		\hline
		LED (mW) & Sample ($\mu$W) & IVC ($\mu$W) & $T$ (mK) & $\dot{Q}$ ($\mu$W)\\ \hline
		0.9 & 0.12 & 1.5 & 9.01 & 0.17 \\
		2.0 & 0.31 & 4.6 & 9.53 & 0.30 \\
		4.0 & 0.50 & 6.7 & 10.21 & 0.47 \\
		9.0 & 1.10 & 15.2 & 11.56 & 0.86 \\
		15.0 & 2.00 & 28.0 & 13.51 & 1.51 \\
		17.5 & 2.50 & 34.5 & 14.45 & 1.85 \\ \hline
	\end{tabular}
\end{table}

\section{Experimental Results}
\label{results}

\begin{figure}
	\includegraphics{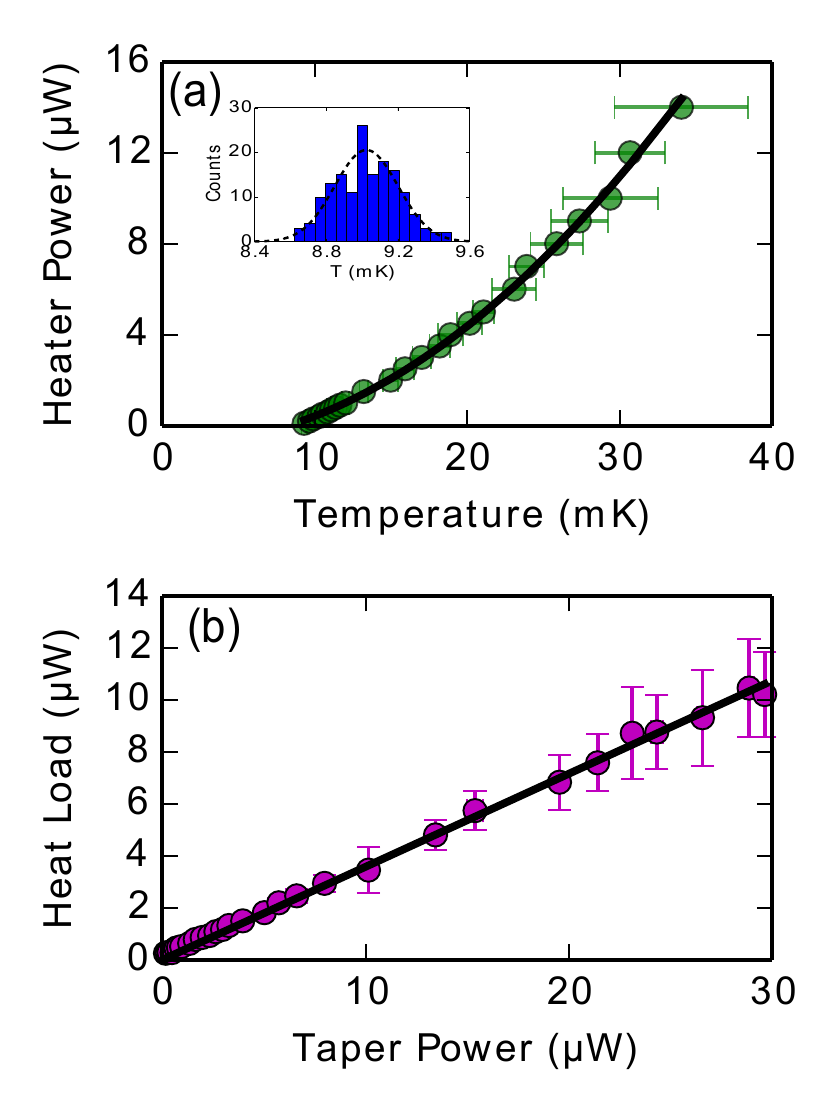}
	\caption{(a) The cooling power of the fridge is measured by applying a fixed heat load to the mixing chamber stage and measuring the resulting temperature. Each data point is an average over at least ten measurements of a nuclear orientation thermometer. The data are fit to a phenomenological model \cite{Pobell2007} (black line) as discussed in the text. Inset: Base temperature of the fridge as measured with the nuclear orientation thermometer. (b) The heat load from laser light in the tapered fiber is inferred from the measured base temperature and (a) as the laser power is increased. The line is a linear fit with a slope of 0.36, indicating that 36\% of the taper power contributes to heating of the mixing chamber. \label{heating}}
\end{figure}

\begin{figure}[b]
	\includegraphics{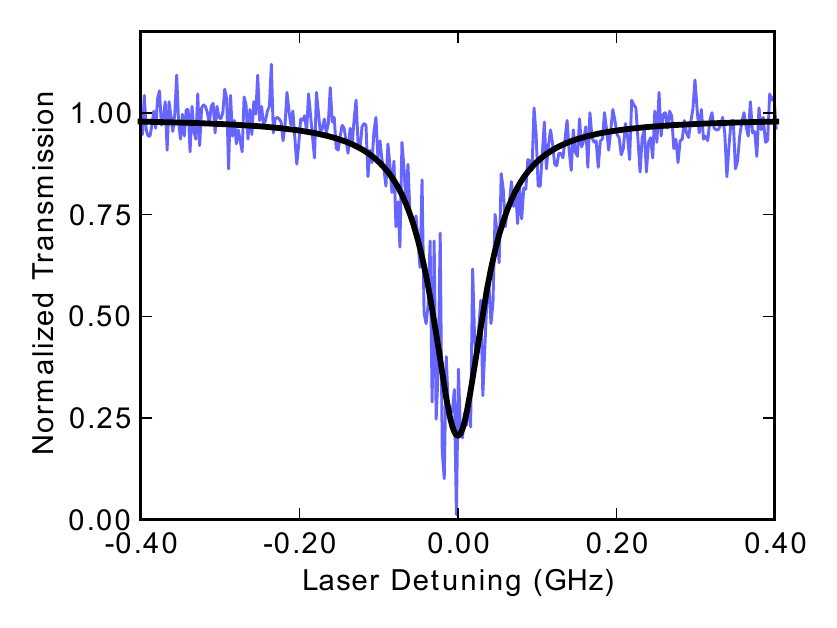}
	\caption{A WGM bottle resonance around 1581 nm with a $Q$ of 2.4$\times10^{6}$, measured at low optical powers ($\sim$250 nW in the tapered region) at the fridge's base temperature of 9 mK.\label{resonance}}
\end{figure}

Our system reaches a base temperature of 9.02\,$\pm$\,0.18 mK (measured by a nuclear orientation primary thermometer and shown in the inset of Figure \ref{heating}) which is only slightly increased over the unloaded fridge base temperature of 7 mK. Investigating potential sources of heating, we find that neither the imaging light nor the motion of the nanopositioning stages cause any perceivable heating under normal experimental conditions. Table \ref{power} tabulates the temperature of the mixing chamber stage and the corresponding heat load for a number of different total green LED powers. To determine this heat load, we first measure the cooling power of the fridge by applying a predetermined heat load $\dot{Q}_{MC}$ to the mixing chamber stage with a resistive heater and measuring the resulting temperature $T$ with the nuclear orientation thermometer (see Figure \ref{heating}). Fitting the measurements over a range of 9-35 mK with $\dot{Q}_{MC}=aT^{2}-\dot{Q}_{HL}$, \cite{Pobell2007} where $a$ is a constant related the $^{3}$He flow rate, we find a constant heat leak $\dot{Q}_{HL}\sim900$ nW. We then infer the LED heat load from the measured base temperature and find that since very little light is required to image the sample, no significant heat load results (see Table \ref{power}). 

The most significant contribution to heating is caused by laser light injected into the tapered measurement fiber. Again inferring the heat load from the base temperature as a function of input power, we find that most of the power lost in the taper ($\sim$36\% of the total power for the taper measured in Figure \ref{heating}(b)) contributes to heating the mixing chamber. For very low optical powers ($<$250 nW in the tapered region), there is no observable increase in temperature, allowing us to probe optical resonances using a 1550 nm tunable external cavity diode laser. By scanning the laser frequency, we find optical resonances with intrinsic quality factors of 10$^{6}$-10$^{7}$, which are consistent with measured room temperature values. An example of an optical resonance at 9 mK is shown in Figure \ref{resonance}.

\begin{figure}[b]
	\includegraphics{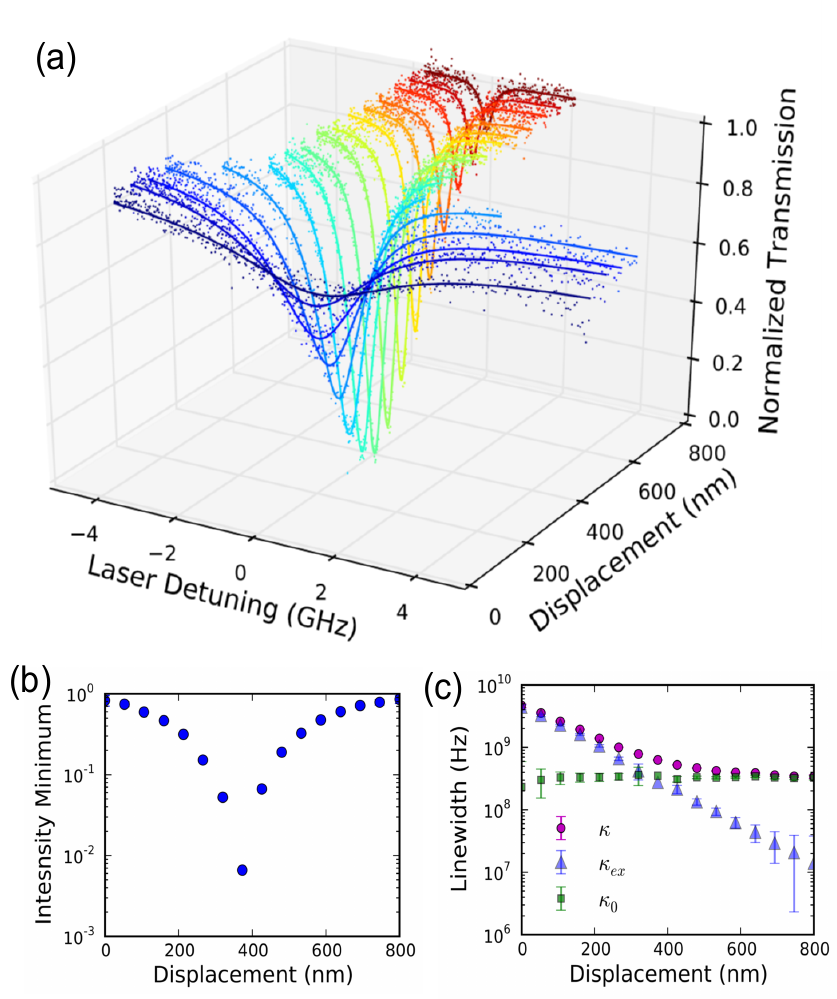}
	\caption{Demonstration of critical coupling at 4.2 K. (a) Transmission across a bottle resonance as a function of displacement of the bottle from the taper. Lines are Lorentzian fits to the data. \cite{Aspelmeyer2013} (b) Measured minima in the normalized transmission, showing an extinction of -22 dB. (c) External ($\kappa_{ex}$), internal ($\kappa_{0}$) and total ($\kappa$) decay rates (linewidths) of the resonance, extracted from the fits in (a). Note the critical coupling point near 375 nm where $\kappa_{0}\approx\kappa_{ex}$. \label{CriticalCoupling}}
\end{figure}

Further investigations of the loaded optical properties at 4.2 K are made by adjusting the continuous voltage offset on the nanopositioning stages, enabling a slow approach of the bottle to the tapered fiber over a total range of 800 nm at 4.2 K. The external coupling rate $\kappa_{ex}$ increases exponentially as the bottle-taper separation is reduced resulting in a transition from under- ($\kappa_{ex}<\kappa_{0}$) to over-coupling ($\kappa_{ex}>\kappa_{0}$) of the resonator. An example of this for an optical resonance around 1598 nm is shown in Figure \ref{CriticalCoupling}. Over-coupling occurs without contacting the resonator (and thus damping mechanical oscillations), an important capability for controlling the mechanical state through the injected light field. \cite{Palomaki2013} We also achieve critical coupling, where the external coupling rate equals the intrinsic decay rate ($\kappa_{ex}=\kappa_{0}$) with a normalized on-resonance transmission minimum of -22 dB. This extinction ratio is a measure of the quality of polarization-, mode- and phase-matching in the system since zero transmission can only occur for perfectly efficient coupling. \cite{Cai2000} Although this is not the first study of cryogenic taper-resonator coupling conditions, \cite{Fujiwara2012} we believe that it is the first to show such complete extinction upon critical coupling. 

\begin{figure}[h]
	\includegraphics{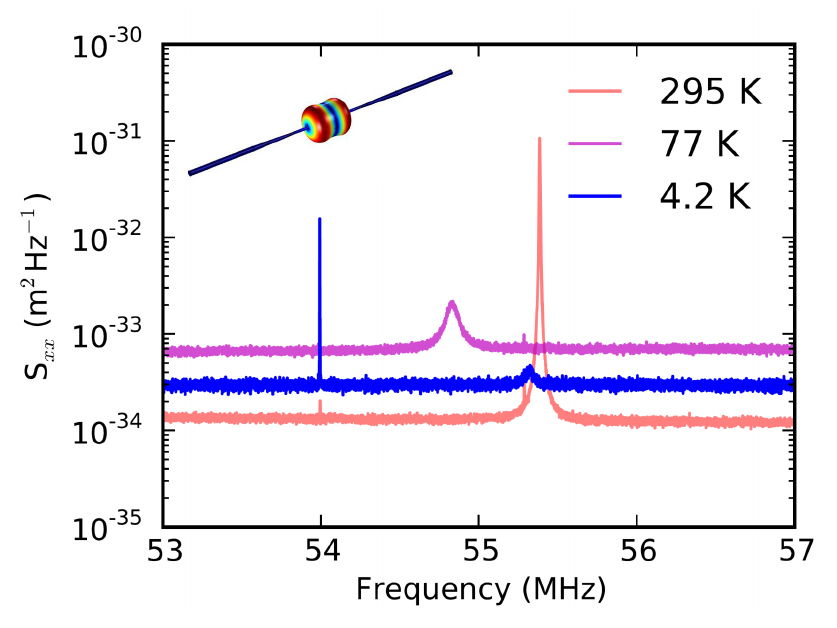}
	\caption{Power spectral density of a 55 MHz breathing mode of a bottle resonator at 295, 77 and 4.2 K, thermalized using exchange gas. The peak at 54 MHz is used to calibrate the temperature of the mechanical mode. \cite{Riviere2013} The inset shows the mode shape, simulated using COMSOL Multiphysics\textsuperscript{TM}.  \label{Mechanics}}
\end{figure}

We finally measure the radial breathing modes of the $\sim$50 $\mu$m bottle resonator by detuning the laser frequency slightly from the optical resonance and recording the fluctuations in the taper transmission that occur as a result of mechanical motion. The high optical quality of the resonators gives rise to a large thermo-optic nonlinearity, \cite{Arcizet2009} leading to heating of the resonator through optical absorption and preventing thermalization of the mechanical modes to base temperature. We instead use helium exchange gas to thermalize the bottles to room, liquid nitrogen and liquid helium temperatures (295, 77 and 4.2 K, respectively). We detect the taper transmission directly on a 1 GHz photodiode and use an analog-to-digital converter to record a time trace of the high-pass filtered signal. Fourier transforming this signal and squaring the result gives direct access to the voltage power spectral density $S_{VV}(f)$ of the detector. Using effective masses determined from finite element modeling, we thermomechanically calibrate \cite{Hauer2013} $S_{VV}(f)$ to obtain the displacement power spectral density $S_{xx}(f)$. The calibrated displacement spectrum of a 55 MHz mechanical breathing mode with a room temperature quality factor of 10$^{4}$ is shown in Figure \ref{Mechanics}.

\section{Conclusion}

We have developed a tapered fiber coupling apparatus on the base plate of a commercial dilution fridge, allowing access to temperatures as low as 9 mK at an overall transmission efficiency of up to 70\%. Our system incorporates three-dimensional control of the resonator-fiber position allowing for full control over the coupling conditions, enabling under-, critical- and over-coupling of the resonator. Furthermore, a coherent bundle of 37,000 optical fibers enables real-time \emph{in situ} imaging of the experiment at a resolution of $\sim$1 $\mu$m with negligible heating. These two properties together facilitate coupling to a large number of either on- or off-chip devices during a single cooldown.

\begin{acknowledgments}
The authors would like to thank Callum Doolin and Hugh Ramp for software support, Fabien Souris for helpful discussions and Paul Zimmerman, Antonio Vinagreiro and Jordan Cameron for technical assistance. This work was supported by the University of Alberta, Faculty of Science, the Natural Sciences and Engineering Research Council of Canada, Alberta Innovates Technology Futures, the Canadian Foundation for Innovation and the Alfred P. Sloan Foundation.
\end{acknowledgments}

\bibliography{EndoscopeBib}

\begin{thebibliography}{46}%
\makeatletter
\providecommand \@ifxundefined [1]{%
 \@ifx{#1\undefined}
}%
\providecommand \@ifnum [1]{%
 \ifnum #1\expandafter \@firstoftwo
 \else \expandafter \@secondoftwo
 \fi
}%
\providecommand \@ifx [1]{%
 \ifx #1\expandafter \@firstoftwo
 \else \expandafter \@secondoftwo
 \fi
}%
\providecommand \natexlab [1]{#1}%
\providecommand \enquote  [1]{``#1''}%
\providecommand \bibnamefont  [1]{#1}%
\providecommand \bibfnamefont [1]{#1}%
\providecommand \citenamefont [1]{#1}%
\providecommand \href@noop [0]{\@secondoftwo}%
\providecommand \href [0]{\begingroup \@sanitize@url \@href}%
\providecommand \@href[1]{\@@startlink{#1}\@@href}%
\providecommand \@@href[1]{\endgroup#1\@@endlink}%
\providecommand \@sanitize@url [0]{\catcode `\\12\catcode `\$12\catcode
  `\&12\catcode `\#12\catcode `\^12\catcode `\_12\catcode `\%12\relax}%
\providecommand \@@startlink[1]{}%
\providecommand \@@endlink[0]{}%
\providecommand \url  [0]{\begingroup\@sanitize@url \@url }%
\providecommand \@url [1]{\endgroup\@href {#1}{\urlprefix }}%
\providecommand \urlprefix  [0]{URL }%
\providecommand \Eprint [0]{\href }%
\providecommand \doibase [0]{http://dx.doi.org/}%
\providecommand \selectlanguage [0]{\@gobble}%
\providecommand \bibinfo  [0]{\@secondoftwo}%
\providecommand \bibfield  [0]{\@secondoftwo}%
\providecommand \translation [1]{[#1]}%
\providecommand \BibitemOpen [0]{}%
\providecommand \bibitemStop [0]{}%
\providecommand \bibitemNoStop [0]{.\EOS\space}%
\providecommand \EOS [0]{\spacefactor3000\relax}%
\providecommand \BibitemShut  [1]{\csname bibitem#1\endcsname}%
\let\auto@bib@innerbib\@empty
\bibitem [{\citenamefont {Aspelmeyer}, \citenamefont {Kippenberg},\ and\
  \citenamefont {Marquardt}(2013)}]{Aspelmeyer2013}%
  \BibitemOpen
  \bibfield  {author} {\bibinfo {author} {\bibfnamefont {M.}~\bibnamefont
  {Aspelmeyer}}, \bibinfo {author} {\bibfnamefont {T.~J.}\ \bibnamefont
  {Kippenberg}}, \ and\ \bibinfo {author} {\bibfnamefont {F.}~\bibnamefont
  {Marquardt}},\ }\href@noop {} {\enquote {\bibinfo {title} {Cavity
  optomechanics},}\ } (\bibinfo {year} {2013}),\ \Eprint
  {http://arxiv.org/abs/arXiv:1303.0733} {arXiv:1303.0733} \BibitemShut
  {NoStop}%
\bibitem [{\citenamefont {Kippenberg}\ and\ \citenamefont
  {Vahala}(2008)}]{Kippenberg2008}%
  \BibitemOpen
  \bibfield  {author} {\bibinfo {author} {\bibfnamefont {T.~J.}\ \bibnamefont
  {Kippenberg}}\ and\ \bibinfo {author} {\bibfnamefont {K.~J.}\ \bibnamefont
  {Vahala}},\ }\href {\doibase 10.1126/science.1156032} {\bibfield  {journal}
  {\bibinfo  {journal} {Science}\ }\textbf {\bibinfo {volume} {321}},\ \bibinfo
  {pages} {1172} (\bibinfo {year} {2008})}\BibitemShut {NoStop}%
\bibitem [{\citenamefont {Gavartin}, \citenamefont {Verlot},\ and\
  \citenamefont {Kippenberg}(2012)}]{Gavartin2012}%
  \BibitemOpen
  \bibfield  {author} {\bibinfo {author} {\bibfnamefont {E.}~\bibnamefont
  {Gavartin}}, \bibinfo {author} {\bibfnamefont {P.}~\bibnamefont {Verlot}}, \
  and\ \bibinfo {author} {\bibfnamefont {T.~J.}\ \bibnamefont {Kippenberg}},\
  }\href {\doibase 10.1038/nnano.2012.97} {\bibfield  {journal} {\bibinfo
  {journal} {Nature Nanotechnology}\ }\textbf {\bibinfo {volume} {7}},\
  \bibinfo {pages} {509} (\bibinfo {year} {2012})}\BibitemShut {NoStop}%
\bibitem [{\citenamefont {Miao}, \citenamefont {Srinivasan},\ and\
  \citenamefont {Aksyuk}(2012)}]{Miao2012}%
  \BibitemOpen
  \bibfield  {author} {\bibinfo {author} {\bibfnamefont {H.}~\bibnamefont
  {Miao}}, \bibinfo {author} {\bibfnamefont {K.}~\bibnamefont {Srinivasan}}, \
  and\ \bibinfo {author} {\bibfnamefont {V.}~\bibnamefont {Aksyuk}},\
  }\href@noop {} {\bibfield  {journal} {\bibinfo  {journal} {New Journal of
  Physics}\ }\textbf {\bibinfo {volume} {14}},\ \bibinfo {pages} {075015}
  (\bibinfo {year} {2012})}\BibitemShut {NoStop}%
\bibitem [{\citenamefont {Doolin}\ \emph {et~al.}(2014)\citenamefont {Doolin},
  \citenamefont {Kim}, \citenamefont {Hauer}, \citenamefont {MacDonald},\ and\
  \citenamefont {Davis}}]{Doolin2014}%
  \BibitemOpen
  \bibfield  {author} {\bibinfo {author} {\bibfnamefont {C.}~\bibnamefont
  {Doolin}}, \bibinfo {author} {\bibfnamefont {P.~H.}\ \bibnamefont {Kim}},
  \bibinfo {author} {\bibfnamefont {B.~D.}\ \bibnamefont {Hauer}}, \bibinfo
  {author} {\bibfnamefont {A.~J.~R.}\ \bibnamefont {MacDonald}}, \ and\
  \bibinfo {author} {\bibfnamefont {J.~P.}\ \bibnamefont {Davis}},\ }\href@noop
  {} {\bibfield  {journal} {\bibinfo  {journal} {New Journal of Physics}\
  }\textbf {\bibinfo {volume} {16}},\ \bibinfo {pages} {035001} (\bibinfo
  {year} {2014})}\BibitemShut {NoStop}%
\bibitem [{\citenamefont {Anetsberger}\ \emph {et~al.}(2010)\citenamefont
  {Anetsberger}, \citenamefont {Gavartin}, \citenamefont {Arcizet},
  \citenamefont {Unterreithmeier}, \citenamefont {Weig}, \citenamefont
  {Gorodetsky}, \citenamefont {Kotthaus},\ and\ \citenamefont
  {Kippenberg}}]{Anetsberger2010}%
  \BibitemOpen
  \bibfield  {author} {\bibinfo {author} {\bibfnamefont {G.}~\bibnamefont
  {Anetsberger}}, \bibinfo {author} {\bibfnamefont {E.}~\bibnamefont
  {Gavartin}}, \bibinfo {author} {\bibfnamefont {O.}~\bibnamefont {Arcizet}},
  \bibinfo {author} {\bibfnamefont {Q.~P.}\ \bibnamefont {Unterreithmeier}},
  \bibinfo {author} {\bibfnamefont {E.~M.}\ \bibnamefont {Weig}}, \bibinfo
  {author} {\bibfnamefont {M.~L.}\ \bibnamefont {Gorodetsky}}, \bibinfo
  {author} {\bibfnamefont {J.~P.}\ \bibnamefont {Kotthaus}}, \ and\ \bibinfo
  {author} {\bibfnamefont {T.~J.}\ \bibnamefont {Kippenberg}},\ }\href
  {\doibase 10.1103/physreva.82.061804} {\bibfield  {journal} {\bibinfo
  {journal} {Physical Review A}\ }\textbf {\bibinfo {volume} {82}},\ \bibinfo
  {pages} {061804(R)} (\bibinfo {year} {2010})}\BibitemShut {NoStop}%
\bibitem [{\citenamefont {Krause}\ \emph {et~al.}(2012)\citenamefont {Krause},
  \citenamefont {Winger}, \citenamefont {Blasius}, \citenamefont {Lin},\ and\
  \citenamefont {Painter}}]{Krause2012}%
  \BibitemOpen
  \bibfield  {author} {\bibinfo {author} {\bibfnamefont {A.~G.}\ \bibnamefont
  {Krause}}, \bibinfo {author} {\bibfnamefont {M.}~\bibnamefont {Winger}},
  \bibinfo {author} {\bibfnamefont {T.~D.}\ \bibnamefont {Blasius}}, \bibinfo
  {author} {\bibfnamefont {Q.}~\bibnamefont {Lin}}, \ and\ \bibinfo {author}
  {\bibfnamefont {O.}~\bibnamefont {Painter}},\ }\href {\doibase
  10.1038/nphoton.2012.245} {\bibfield  {journal} {\bibinfo  {journal} {Nature
  Photonics}\ }\textbf {\bibinfo {volume} {6}},\ \bibinfo {pages} {768}
  (\bibinfo {year} {2012})}\BibitemShut {NoStop}%
\bibitem [{\citenamefont {Kim}\ \emph {et~al.}(2013)\citenamefont {Kim},
  \citenamefont {Doolin}, \citenamefont {Hauer}, \citenamefont {MacDonald},
  \citenamefont {Freeman}, \citenamefont {Barclay},\ and\ \citenamefont
  {Davis}}]{Kim2013}%
  \BibitemOpen
  \bibfield  {author} {\bibinfo {author} {\bibfnamefont {P.~H.}\ \bibnamefont
  {Kim}}, \bibinfo {author} {\bibfnamefont {C.}~\bibnamefont {Doolin}},
  \bibinfo {author} {\bibfnamefont {B.~D.}\ \bibnamefont {Hauer}}, \bibinfo
  {author} {\bibfnamefont {A.~J.~R.}\ \bibnamefont {MacDonald}}, \bibinfo
  {author} {\bibfnamefont {M.~R.}\ \bibnamefont {Freeman}}, \bibinfo {author}
  {\bibfnamefont {P.~E.}\ \bibnamefont {Barclay}}, \ and\ \bibinfo {author}
  {\bibfnamefont {J.~P.}\ \bibnamefont {Davis}},\ }\href {\doibase
  10.1063/1.4789442} {\bibfield  {journal} {\bibinfo  {journal} {Applied
  Physics Letters}\ }\textbf {\bibinfo {volume} {102}},\ \bibinfo {pages}
  {053102} (\bibinfo {year} {2013})}\BibitemShut {NoStop}%
\bibitem [{\citenamefont {Wu}\ \emph {et~al.}(2014)\citenamefont {Wu},
  \citenamefont {Hryciw}, \citenamefont {Healey}, \citenamefont {Lake},
  \citenamefont {Jayakumar}, \citenamefont {Freeman}, \citenamefont {Davis},\
  and\ \citenamefont {Barclay}}]{Wu2014}%
  \BibitemOpen
  \bibfield  {author} {\bibinfo {author} {\bibfnamefont {M.}~\bibnamefont
  {Wu}}, \bibinfo {author} {\bibfnamefont {A.~C.}\ \bibnamefont {Hryciw}},
  \bibinfo {author} {\bibfnamefont {C.}~\bibnamefont {Healey}}, \bibinfo
  {author} {\bibfnamefont {D.~P.}\ \bibnamefont {Lake}}, \bibinfo {author}
  {\bibfnamefont {H.}~\bibnamefont {Jayakumar}}, \bibinfo {author}
  {\bibfnamefont {M.~R.}\ \bibnamefont {Freeman}}, \bibinfo {author}
  {\bibfnamefont {J.~P.}\ \bibnamefont {Davis}}, \ and\ \bibinfo {author}
  {\bibfnamefont {P.~E.}\ \bibnamefont {Barclay}},\ }\href {\doibase
  10.1103/physrevx.4.021052} {\bibfield  {journal} {\bibinfo  {journal}
  {Physical Review X}\ }\textbf {\bibinfo {volume} {4}},\ \bibinfo {pages}
  {021052} (\bibinfo {year} {2014})}\BibitemShut {NoStop}%
\bibitem [{\citenamefont {Stannigel}\ \emph {et~al.}(2010)\citenamefont
  {Stannigel}, \citenamefont {Rabl}, \citenamefont {S$\o$rensen}, \citenamefont
  {Zoller},\ and\ \citenamefont {Lukin}}]{Stannigel2010}%
  \BibitemOpen
  \bibfield  {author} {\bibinfo {author} {\bibfnamefont {K.}~\bibnamefont
  {Stannigel}}, \bibinfo {author} {\bibfnamefont {P.}~\bibnamefont {Rabl}},
  \bibinfo {author} {\bibfnamefont {A.~S.}\ \bibnamefont {S$\o$rensen}},
  \bibinfo {author} {\bibfnamefont {P.}~\bibnamefont {Zoller}}, \ and\ \bibinfo
  {author} {\bibfnamefont {M.~D.}\ \bibnamefont {Lukin}},\ }\href {\doibase
  10.1103/physrevlett.105.220501} {\bibfield  {journal} {\bibinfo  {journal}
  {Physical Review Letters}\ }\textbf {\bibinfo {volume} {105}},\ \bibinfo
  {pages} {220501} (\bibinfo {year} {2010})}\BibitemShut {NoStop}%
\bibitem [{\citenamefont {O'Connell}\ \emph {et~al.}(2010)\citenamefont
  {O'Connell}, \citenamefont {Hofheinz}, \citenamefont {Ansmann}, \citenamefont
  {Bialczak}, \citenamefont {Lenander}, \citenamefont {Lucero}, \citenamefont
  {Neeley}, \citenamefont {Sank}, \citenamefont {Wang}, \citenamefont {Weides},
  \citenamefont {Wenner}, \citenamefont {Martinis},\ and\ \citenamefont
  {Cleland}}]{OConnell2010}%
  \BibitemOpen
  \bibfield  {author} {\bibinfo {author} {\bibfnamefont {A.~D.}\ \bibnamefont
  {O'Connell}}, \bibinfo {author} {\bibfnamefont {M.}~\bibnamefont {Hofheinz}},
  \bibinfo {author} {\bibfnamefont {M.}~\bibnamefont {Ansmann}}, \bibinfo
  {author} {\bibfnamefont {R.~C.}\ \bibnamefont {Bialczak}}, \bibinfo {author}
  {\bibfnamefont {M.}~\bibnamefont {Lenander}}, \bibinfo {author}
  {\bibfnamefont {E.}~\bibnamefont {Lucero}}, \bibinfo {author} {\bibfnamefont
  {M.}~\bibnamefont {Neeley}}, \bibinfo {author} {\bibfnamefont
  {D.}~\bibnamefont {Sank}}, \bibinfo {author} {\bibfnamefont {H.}~\bibnamefont
  {Wang}}, \bibinfo {author} {\bibfnamefont {M.}~\bibnamefont {Weides}},
  \bibinfo {author} {\bibfnamefont {J.}~\bibnamefont {Wenner}}, \bibinfo
  {author} {\bibfnamefont {J.~M.}\ \bibnamefont {Martinis}}, \ and\ \bibinfo
  {author} {\bibfnamefont {A.~N.}\ \bibnamefont {Cleland}},\ }\href {\doibase
  10.1038/nature08967} {\bibfield  {journal} {\bibinfo  {journal} {Nature}\
  }\textbf {\bibinfo {volume} {464}},\ \bibinfo {pages} {697} (\bibinfo {year}
  {2010})}\BibitemShut {NoStop}%
\bibitem [{\citenamefont {Bochmann}\ \emph {et~al.}(2013)\citenamefont
  {Bochmann}, \citenamefont {Vainsencher}, \citenamefont {Awschalom},\ and\
  \citenamefont {Cleland}}]{Bochmann2013}%
  \BibitemOpen
  \bibfield  {author} {\bibinfo {author} {\bibfnamefont {J.}~\bibnamefont
  {Bochmann}}, \bibinfo {author} {\bibfnamefont {A.}~\bibnamefont
  {Vainsencher}}, \bibinfo {author} {\bibfnamefont {D.~D.}\ \bibnamefont
  {Awschalom}}, \ and\ \bibinfo {author} {\bibfnamefont {A.~N.}\ \bibnamefont
  {Cleland}},\ }\href@noop {} {\bibfield  {journal} {\bibinfo  {journal}
  {Nature Physics}\ }\textbf {\bibinfo {volume} {9}},\ \bibinfo {pages} {712}
  (\bibinfo {year} {2013})}\BibitemShut {NoStop}%
\bibitem [{\citenamefont {Safavi-Naeini}\ \emph {et~al.}(2013)\citenamefont
  {Safavi-Naeini}, \citenamefont {Gr\"{o}blacher}, \citenamefont {Hill},
  \citenamefont {Chan}, \citenamefont {Aspelmeyer},\ and\ \citenamefont
  {Painter}}]{SafaviNaeini2013}%
  \BibitemOpen
  \bibfield  {author} {\bibinfo {author} {\bibfnamefont {A.~H.}\ \bibnamefont
  {Safavi-Naeini}}, \bibinfo {author} {\bibfnamefont {S.}~\bibnamefont
  {Gr\"{o}blacher}}, \bibinfo {author} {\bibfnamefont {J.~T.}\ \bibnamefont
  {Hill}}, \bibinfo {author} {\bibfnamefont {J.}~\bibnamefont {Chan}}, \bibinfo
  {author} {\bibfnamefont {M.}~\bibnamefont {Aspelmeyer}}, \ and\ \bibinfo
  {author} {\bibfnamefont {O.}~\bibnamefont {Painter}},\ }\href@noop {}
  {\bibfield  {journal} {\bibinfo  {journal} {Nature}\ }\textbf {\bibinfo
  {volume} {500}},\ \bibinfo {pages} {185} (\bibinfo {year}
  {2013})}\BibitemShut {NoStop}%
\bibitem [{\citenamefont {Palomaki}\ \emph {et~al.}(2013)\citenamefont
  {Palomaki}, \citenamefont {Teufel}, \citenamefont {Simmonds},\ and\
  \citenamefont {Lehnert}}]{Palomaki2013}%
  \BibitemOpen
  \bibfield  {author} {\bibinfo {author} {\bibfnamefont {T.~A.}\ \bibnamefont
  {Palomaki}}, \bibinfo {author} {\bibfnamefont {J.~D.}\ \bibnamefont
  {Teufel}}, \bibinfo {author} {\bibfnamefont {R.~W.}\ \bibnamefont
  {Simmonds}}, \ and\ \bibinfo {author} {\bibfnamefont {K.~W.}\ \bibnamefont
  {Lehnert}},\ }\href {\doibase 10.1126/science.1244563} {\bibfield  {journal}
  {\bibinfo  {journal} {Science}\ }\textbf {\bibinfo {volume} {342}},\ \bibinfo
  {pages} {710} (\bibinfo {year} {2013})}\BibitemShut {NoStop}%
\bibitem [{\citenamefont {B$\o$rkje}(2014)}]{Borkje2014}%
  \BibitemOpen
  \bibfield  {author} {\bibinfo {author} {\bibfnamefont {K.}~\bibnamefont
  {B$\o$rkje}},\ }\href {\doibase 10.1103/physreva.90.023806} {\bibfield
  {journal} {\bibinfo  {journal} {Physical Review A}\ }\textbf {\bibinfo
  {volume} {90}},\ \bibinfo {pages} {023806} (\bibinfo {year}
  {2014})}\BibitemShut {NoStop}%
\bibitem [{\citenamefont {Schliesser}\ \emph {et~al.}(2008)\citenamefont
  {Schliesser}, \citenamefont {Rivi\`{e}re}, \citenamefont {Anetsberger},
  \citenamefont {Arcizet},\ and\ \citenamefont {Kippenberg}}]{Schliesser2008}%
  \BibitemOpen
  \bibfield  {author} {\bibinfo {author} {\bibfnamefont {A.}~\bibnamefont
  {Schliesser}}, \bibinfo {author} {\bibfnamefont {R.}~\bibnamefont
  {Rivi\`{e}re}}, \bibinfo {author} {\bibfnamefont {G.}~\bibnamefont
  {Anetsberger}}, \bibinfo {author} {\bibfnamefont {O.}~\bibnamefont
  {Arcizet}}, \ and\ \bibinfo {author} {\bibfnamefont {T.~J.}\ \bibnamefont
  {Kippenberg}},\ }\href {\doibase 10.1038/nphys939} {\bibfield  {journal}
  {\bibinfo  {journal} {Nature Physics}\ }\textbf {\bibinfo {volume} {4}},\
  \bibinfo {pages} {415} (\bibinfo {year} {2008})}\BibitemShut {NoStop}%
\bibitem [{\citenamefont {Chan}\ \emph {et~al.}(2011)\citenamefont {Chan},
  \citenamefont {Mayer~Alegre}, \citenamefont {Safavi-Naeini}, \citenamefont
  {Hill}, \citenamefont {Krause}, \citenamefont {Gr\"{o}blacher}, \citenamefont
  {Aspelmeyer},\ and\ \citenamefont {Painter}}]{Chan2011}%
  \BibitemOpen
  \bibfield  {author} {\bibinfo {author} {\bibfnamefont {J.}~\bibnamefont
  {Chan}}, \bibinfo {author} {\bibfnamefont {T.~P.}\ \bibnamefont
  {Mayer~Alegre}}, \bibinfo {author} {\bibfnamefont {A.~H.}\ \bibnamefont
  {Safavi-Naeini}}, \bibinfo {author} {\bibfnamefont {J.~T.}\ \bibnamefont
  {Hill}}, \bibinfo {author} {\bibfnamefont {A.}~\bibnamefont {Krause}},
  \bibinfo {author} {\bibfnamefont {S.}~\bibnamefont {Gr\"{o}blacher}},
  \bibinfo {author} {\bibfnamefont {M.}~\bibnamefont {Aspelmeyer}}, \ and\
  \bibinfo {author} {\bibfnamefont {O.}~\bibnamefont {Painter}},\ }\href
  {\doibase 10.1038/nature10461} {\bibfield  {journal} {\bibinfo  {journal}
  {Nature}\ }\textbf {\bibinfo {volume} {478}},\ \bibinfo {pages} {89}
  (\bibinfo {year} {2011})}\BibitemShut {NoStop}%
\bibitem [{\citenamefont {Teufel}\ \emph {et~al.}(2011)\citenamefont {Teufel},
  \citenamefont {Donner}, \citenamefont {Li}, \citenamefont {Harlow},
  \citenamefont {Allman}, \citenamefont {Cicak}, \citenamefont {Sirois},
  \citenamefont {Whittaker}, \citenamefont {Lehnert},\ and\ \citenamefont
  {Simmonds}}]{Teufel2011}%
  \BibitemOpen
  \bibfield  {author} {\bibinfo {author} {\bibfnamefont {J.~D.}\ \bibnamefont
  {Teufel}}, \bibinfo {author} {\bibfnamefont {T.}~\bibnamefont {Donner}},
  \bibinfo {author} {\bibfnamefont {D.}~\bibnamefont {Li}}, \bibinfo {author}
  {\bibfnamefont {J.~W.}\ \bibnamefont {Harlow}}, \bibinfo {author}
  {\bibfnamefont {M.~S.}\ \bibnamefont {Allman}}, \bibinfo {author}
  {\bibfnamefont {K.}~\bibnamefont {Cicak}}, \bibinfo {author} {\bibfnamefont
  {A.~J.}\ \bibnamefont {Sirois}}, \bibinfo {author} {\bibfnamefont {J.~D.}\
  \bibnamefont {Whittaker}}, \bibinfo {author} {\bibfnamefont {K.~W.}\
  \bibnamefont {Lehnert}}, \ and\ \bibinfo {author} {\bibfnamefont {R.~W.}\
  \bibnamefont {Simmonds}},\ }\href {\doibase 10.1038/nature10261} {\bibfield
  {journal} {\bibinfo  {journal} {Nature}\ }\textbf {\bibinfo {volume} {475}},\
  \bibinfo {pages} {359} (\bibinfo {year} {2011})}\BibitemShut {NoStop}%
\bibitem [{\citenamefont {Park}\ and\ \citenamefont {Wang}(2009)}]{Park2009}%
  \BibitemOpen
  \bibfield  {author} {\bibinfo {author} {\bibfnamefont {Y.-S.}\ \bibnamefont
  {Park}}\ and\ \bibinfo {author} {\bibfnamefont {H.}~\bibnamefont {Wang}},\
  }\href {\doibase 10.1038/nphys1303} {\bibfield  {journal} {\bibinfo
  {journal} {Nature Physics}\ }\textbf {\bibinfo {volume} {5}},\ \bibinfo
  {pages} {489} (\bibinfo {year} {2009})}\BibitemShut {NoStop}%
\bibitem [{\citenamefont {Sun}\ \emph {et~al.}(2013)\citenamefont {Sun},
  \citenamefont {Zhang}, \citenamefont {Schuck},\ and\ \citenamefont
  {Tang}}]{Sun13}%
  \BibitemOpen
  \bibfield  {author} {\bibinfo {author} {\bibfnamefont {X.}~\bibnamefont
  {Sun}}, \bibinfo {author} {\bibfnamefont {X.}~\bibnamefont {Zhang}}, \bibinfo
  {author} {\bibfnamefont {C.}~\bibnamefont {Schuck}}, \ and\ \bibinfo {author}
  {\bibfnamefont {H.~X.}\ \bibnamefont {Tang}},\ }\href {\doibase
  10.1038/srep01436} {\bibfield  {journal} {\bibinfo  {journal} {Scientific
  Reports}\ }\textbf {\bibinfo {volume} {3}},\ \bibinfo {pages} {1436}
  (\bibinfo {year} {2013})}\BibitemShut {NoStop}%
\bibitem [{\citenamefont {De~Lorenzo}\ and\ \citenamefont
  {Schwab}(2014)}]{DeLorenzo2014}%
  \BibitemOpen
  \bibfield  {author} {\bibinfo {author} {\bibfnamefont {L.~A.}\ \bibnamefont
  {De~Lorenzo}}\ and\ \bibinfo {author} {\bibfnamefont {K.~C.}\ \bibnamefont
  {Schwab}},\ }\href@noop {} {\bibfield  {journal} {\bibinfo  {journal} {New
  Journal of Physics}\ }\textbf {\bibinfo {volume} {16}},\ \bibinfo {pages}
  {113020} (\bibinfo {year} {2014})}\BibitemShut {NoStop}%
\bibitem [{\citenamefont {Srinivasan}\ and\ \citenamefont
  {Painter}(2007)}]{Srinivasan2007}%
  \BibitemOpen
  \bibfield  {author} {\bibinfo {author} {\bibfnamefont {K.}~\bibnamefont
  {Srinivasan}}\ and\ \bibinfo {author} {\bibfnamefont {O.}~\bibnamefont
  {Painter}},\ }\href {\doibase 10.1063/1.2431719} {\bibfield  {journal}
  {\bibinfo  {journal} {Applied Physics Letters}\ }\textbf {\bibinfo {volume}
  {90}},\ \bibinfo {pages} {031114} (\bibinfo {year} {2007})}\BibitemShut
  {NoStop}%
\bibitem [{\citenamefont {Rivi\`{e}re}\ \emph {et~al.}(2013)\citenamefont
  {Rivi\`{e}re}, \citenamefont {Arcizet}, \citenamefont {Schliesser},\ and\
  \citenamefont {Kippenberg}}]{Riviere2013}%
  \BibitemOpen
  \bibfield  {author} {\bibinfo {author} {\bibfnamefont {R.}~\bibnamefont
  {Rivi\`{e}re}}, \bibinfo {author} {\bibfnamefont {O.}~\bibnamefont
  {Arcizet}}, \bibinfo {author} {\bibfnamefont {A.}~\bibnamefont {Schliesser}},
  \ and\ \bibinfo {author} {\bibfnamefont {T.~J.}\ \bibnamefont {Kippenberg}},\
  }\href {\doibase 10.1063/1.4801456} {\bibfield  {journal} {\bibinfo
  {journal} {Review of Scientific Instruments}\ }\textbf {\bibinfo {volume}
  {84}},\ \bibinfo {pages} {043108} (\bibinfo {year} {2013})}\BibitemShut
  {NoStop}%
\bibitem [{\citenamefont {Meenehan}\ \emph {et~al.}(2014)\citenamefont
  {Meenehan}, \citenamefont {Cohen}, \citenamefont {Gr\"{o}blacher},
  \citenamefont {Hill}, \citenamefont {Safavi-Naeini}, \citenamefont
  {Aspelmeyer},\ and\ \citenamefont {Painter}}]{Meenehan2014}%
  \BibitemOpen
  \bibfield  {author} {\bibinfo {author} {\bibfnamefont {S.~M.}\ \bibnamefont
  {Meenehan}}, \bibinfo {author} {\bibfnamefont {J.~D.}\ \bibnamefont {Cohen}},
  \bibinfo {author} {\bibfnamefont {S.}~\bibnamefont {Gr\"{o}blacher}},
  \bibinfo {author} {\bibfnamefont {J.~T.}\ \bibnamefont {Hill}}, \bibinfo
  {author} {\bibfnamefont {A.~H.}\ \bibnamefont {Safavi-Naeini}}, \bibinfo
  {author} {\bibfnamefont {M.}~\bibnamefont {Aspelmeyer}}, \ and\ \bibinfo
  {author} {\bibfnamefont {O.}~\bibnamefont {Painter}},\ }\href {\doibase
  10.1103/physreva.90.011803} {\bibfield  {journal} {\bibinfo  {journal}
  {Physical Review A}\ }\textbf {\bibinfo {volume} {90}},\ \bibinfo {pages}
  {011803(R)} (\bibinfo {year} {2014})}\BibitemShut {NoStop}%
\bibitem [{\citenamefont {Hauer}\ \emph {et~al.}(2014)\citenamefont {Hauer},
  \citenamefont {Kim}, \citenamefont {Doolin}, \citenamefont {MacDonald},
  \citenamefont {Ramp},\ and\ \citenamefont {Davis}}]{Hauer2014}%
  \BibitemOpen
  \bibfield  {author} {\bibinfo {author} {\bibfnamefont {B.~D.}\ \bibnamefont
  {Hauer}}, \bibinfo {author} {\bibfnamefont {P.~H.}\ \bibnamefont {Kim}},
  \bibinfo {author} {\bibfnamefont {C.}~\bibnamefont {Doolin}}, \bibinfo
  {author} {\bibfnamefont {A.~J.~R.}\ \bibnamefont {MacDonald}}, \bibinfo
  {author} {\bibfnamefont {H.}~\bibnamefont {Ramp}}, \ and\ \bibinfo {author}
  {\bibfnamefont {J.~P.}\ \bibnamefont {Davis}},\ }\href@noop {} {\bibfield
  {journal} {\bibinfo  {journal} {EPJ Techniques and Instrumentation}\ }\textbf
  {\bibinfo {volume} {1:4}} (\bibinfo {year} {2014})}\BibitemShut {NoStop}%
\bibitem [{\citenamefont {Kakarantzas}\ \emph {et~al.}(2001)\citenamefont
  {Kakarantzas}, \citenamefont {Dimmick}, \citenamefont {Birks}, \citenamefont
  {Le~Roux},\ and\ \citenamefont {Russell}}]{Kakarantzas2001}%
  \BibitemOpen
  \bibfield  {author} {\bibinfo {author} {\bibfnamefont {G.}~\bibnamefont
  {Kakarantzas}}, \bibinfo {author} {\bibfnamefont {T.~E.}\ \bibnamefont
  {Dimmick}}, \bibinfo {author} {\bibfnamefont {T.~A.}\ \bibnamefont {Birks}},
  \bibinfo {author} {\bibfnamefont {R.}~\bibnamefont {Le~Roux}}, \ and\
  \bibinfo {author} {\bibfnamefont {P.~S.~J.}\ \bibnamefont {Russell}},\
  }\href@noop {} {\bibfield  {journal} {\bibinfo  {journal} {Optics Letters}\
  }\textbf {\bibinfo {volume} {26}},\ \bibinfo {pages} {1137} (\bibinfo {year}
  {2001})}\BibitemShut {NoStop}%
\bibitem [{\citenamefont {P\"{o}llinger}\ \emph {et~al.}(2009)\citenamefont
  {P\"{o}llinger}, \citenamefont {O'Shea}, \citenamefont {Warken},\ and\
  \citenamefont {Rauschenbeutel}}]{Pollinger2009}%
  \BibitemOpen
  \bibfield  {author} {\bibinfo {author} {\bibfnamefont {M.}~\bibnamefont
  {P\"{o}llinger}}, \bibinfo {author} {\bibfnamefont {D.}~\bibnamefont
  {O'Shea}}, \bibinfo {author} {\bibfnamefont {F.}~\bibnamefont {Warken}}, \
  and\ \bibinfo {author} {\bibfnamefont {A.}~\bibnamefont {Rauschenbeutel}},\
  }\href {\doibase 10.1103/physrevlett.103.053901} {\bibfield  {journal}
  {\bibinfo  {journal} {Physical Review Letters}\ }\textbf {\bibinfo {volume}
  {103}},\ \bibinfo {pages} {053901} (\bibinfo {year} {2009})}\BibitemShut
  {NoStop}%
\bibitem [{\citenamefont {Knight}\ \emph {et~al.}(1997)\citenamefont {Knight},
  \citenamefont {Cheung}, \citenamefont {Jacques},\ and\ \citenamefont
  {Birks}}]{Knight1997}%
  \BibitemOpen
  \bibfield  {author} {\bibinfo {author} {\bibfnamefont {J.~C.}\ \bibnamefont
  {Knight}}, \bibinfo {author} {\bibfnamefont {G.}~\bibnamefont {Cheung}},
  \bibinfo {author} {\bibfnamefont {F.}~\bibnamefont {Jacques}}, \ and\
  \bibinfo {author} {\bibfnamefont {T.~A.}\ \bibnamefont {Birks}},\ }\href@noop
  {} {\bibfield  {journal} {\bibinfo  {journal} {Optics Letters}\ }\textbf
  {\bibinfo {volume} {22}},\ \bibinfo {pages} {1129} (\bibinfo {year}
  {1997})}\BibitemShut {NoStop}%
\bibitem [{\citenamefont {Cai}, \citenamefont {Painter},\ and\ \citenamefont
  {Vahala}(2000)}]{Cai2000}%
  \BibitemOpen
  \bibfield  {author} {\bibinfo {author} {\bibfnamefont {M.}~\bibnamefont
  {Cai}}, \bibinfo {author} {\bibfnamefont {O.}~\bibnamefont {Painter}}, \ and\
  \bibinfo {author} {\bibfnamefont {K.~J.}\ \bibnamefont {Vahala}},\
  }\href@noop {} {\bibfield  {journal} {\bibinfo  {journal} {Physical Review
  Letters}\ }\textbf {\bibinfo {volume} {85}},\ \bibinfo {pages} {74} (\bibinfo
  {year} {2000})}\BibitemShut {NoStop}%
\bibitem [{\citenamefont {Srinivasan}\ \emph {et~al.}(2005)\citenamefont
  {Srinivasan}, \citenamefont {Stintz}, \citenamefont {Krishna},\ and\
  \citenamefont {Painter}}]{Srinivasan2005}%
  \BibitemOpen
  \bibfield  {author} {\bibinfo {author} {\bibfnamefont {K.}~\bibnamefont
  {Srinivasan}}, \bibinfo {author} {\bibfnamefont {A.}~\bibnamefont {Stintz}},
  \bibinfo {author} {\bibfnamefont {S.}~\bibnamefont {Krishna}}, \ and\
  \bibinfo {author} {\bibfnamefont {O.}~\bibnamefont {Painter}},\ }\href
  {\doibase 10.1103/physrevb.72.205318} {\bibfield  {journal} {\bibinfo
  {journal} {Physical Review B}\ }\textbf {\bibinfo {volume} {72}},\ \bibinfo
  {pages} {205318} (\bibinfo {year} {2005})}\BibitemShut {NoStop}%
\bibitem [{\citenamefont {Park}\ and\ \citenamefont {Wang}(2007)}]{Park2007}%
  \BibitemOpen
  \bibfield  {author} {\bibinfo {author} {\bibfnamefont {Y.-S.}\ \bibnamefont
  {Park}}\ and\ \bibinfo {author} {\bibfnamefont {H.}~\bibnamefont {Wang}},\
  }\href@noop {} {\bibfield  {journal} {\bibinfo  {journal} {Optics Letters}\
  }\textbf {\bibinfo {volume} {32}},\ \bibinfo {pages} {3104} (\bibinfo {year}
  {2007})}\BibitemShut {NoStop}%
\bibitem [{\citenamefont {Treussart}\ \emph {et~al.}(1998)\citenamefont
  {Treussart}, \citenamefont {Ilchenko}, \citenamefont {Roch}, \citenamefont
  {Hare}, \citenamefont {Lef\`{e}vre-Seguin}, \citenamefont {Raimond},\ and\
  \citenamefont {Haroche}}]{Treussart1998}%
  \BibitemOpen
  \bibfield  {author} {\bibinfo {author} {\bibfnamefont {F.}~\bibnamefont
  {Treussart}}, \bibinfo {author} {\bibfnamefont {V.~S.}\ \bibnamefont
  {Ilchenko}}, \bibinfo {author} {\bibfnamefont {J.-F.}\ \bibnamefont {Roch}},
  \bibinfo {author} {\bibfnamefont {J.}~\bibnamefont {Hare}}, \bibinfo {author}
  {\bibfnamefont {V.}~\bibnamefont {Lef\`{e}vre-Seguin}}, \bibinfo {author}
  {\bibfnamefont {J.-M.}\ \bibnamefont {Raimond}}, \ and\ \bibinfo {author}
  {\bibfnamefont {S.}~\bibnamefont {Haroche}},\ }\href@noop {} {\bibfield
  {journal} {\bibinfo  {journal} {European Physcal Journal D}\ }\textbf
  {\bibinfo {volume} {1}},\ \bibinfo {pages} {235} (\bibinfo {year}
  {1998})}\BibitemShut {NoStop}%
\bibitem [{\citenamefont {Takashima}\ \emph {et~al.}(2010)\citenamefont
  {Takashima}, \citenamefont {Asai}, \citenamefont {Toubaru}, \citenamefont
  {Fujiwara}, \citenamefont {Sasaki},\ and\ \citenamefont
  {Takeuchi}}]{Takashima2010}%
  \BibitemOpen
  \bibfield  {author} {\bibinfo {author} {\bibfnamefont {H.}~\bibnamefont
  {Takashima}}, \bibinfo {author} {\bibfnamefont {T.}~\bibnamefont {Asai}},
  \bibinfo {author} {\bibfnamefont {K.}~\bibnamefont {Toubaru}}, \bibinfo
  {author} {\bibfnamefont {M.}~\bibnamefont {Fujiwara}}, \bibinfo {author}
  {\bibfnamefont {K.}~\bibnamefont {Sasaki}}, \ and\ \bibinfo {author}
  {\bibfnamefont {S.}~\bibnamefont {Takeuchi}},\ }\href@noop {} {\bibfield
  {journal} {\bibinfo  {journal} {Optics Express}\ }\textbf {\bibinfo {volume}
  {18}},\ \bibinfo {pages} {15169} (\bibinfo {year} {2010})}\BibitemShut
  {NoStop}%
\bibitem [{\citenamefont {Balibar}, \citenamefont {Alles},\ and\ \citenamefont
  {Parshin}(2005)}]{Balibar2005}%
  \BibitemOpen
  \bibfield  {author} {\bibinfo {author} {\bibfnamefont {S.}~\bibnamefont
  {Balibar}}, \bibinfo {author} {\bibfnamefont {H.}~\bibnamefont {Alles}}, \
  and\ \bibinfo {author} {\bibfnamefont {A.~Y.}\ \bibnamefont {Parshin}},\
  }\href@noop {} {\bibfield  {journal} {\bibinfo  {journal} {Reviews of Modern
  Physics}\ }\textbf {\bibinfo {volume} {77}},\ \bibinfo {pages} {317}
  (\bibinfo {year} {2005})}\BibitemShut {NoStop}%
\bibitem [{\citenamefont {Franck}\ and\ \citenamefont
  {Jung}(1986)}]{Franck1986}%
  \BibitemOpen
  \bibfield  {author} {\bibinfo {author} {\bibfnamefont {J.~P.}\ \bibnamefont
  {Franck}}\ and\ \bibinfo {author} {\bibfnamefont {J.}~\bibnamefont {Jung}},\
  }\href@noop {} {\bibfield  {journal} {\bibinfo  {journal} {Journal of Low
  Temperature Physics}\ }\textbf {\bibinfo {volume} {64}},\ \bibinfo {pages}
  {165} (\bibinfo {year} {1986})}\BibitemShut {NoStop}%
\bibitem [{\citenamefont {Nomura}\ \emph {et~al.}(2003)\citenamefont {Nomura},
  \citenamefont {Suzuki}, \citenamefont {Kimura},\ and\ \citenamefont
  {Okuda}}]{Nomura2003}%
  \BibitemOpen
  \bibfield  {author} {\bibinfo {author} {\bibfnamefont {R.}~\bibnamefont
  {Nomura}}, \bibinfo {author} {\bibfnamefont {Y.}~\bibnamefont {Suzuki}},
  \bibinfo {author} {\bibfnamefont {S.}~\bibnamefont {Kimura}}, \ and\ \bibinfo
  {author} {\bibfnamefont {Y.}~\bibnamefont {Okuda}},\ }\href@noop {}
  {\bibfield  {journal} {\bibinfo  {journal} {Physical Review Letters}\
  }\textbf {\bibinfo {volume} {90}},\ \bibinfo {pages} {075301} (\bibinfo
  {year} {2003})}\BibitemShut {NoStop}%
\bibitem [{\citenamefont {Rojas}\ \emph {et~al.}(2010)\citenamefont {Rojas},
  \citenamefont {Haziot}, \citenamefont {Bapst}, \citenamefont {Balibar},\ and\
  \citenamefont {Maris}}]{Rojas2010}%
  \BibitemOpen
  \bibfield  {author} {\bibinfo {author} {\bibfnamefont {X.}~\bibnamefont
  {Rojas}}, \bibinfo {author} {\bibfnamefont {A.}~\bibnamefont {Haziot}},
  \bibinfo {author} {\bibfnamefont {V.}~\bibnamefont {Bapst}}, \bibinfo
  {author} {\bibfnamefont {S.}~\bibnamefont {Balibar}}, \ and\ \bibinfo
  {author} {\bibfnamefont {H.~J.}\ \bibnamefont {Maris}},\ }\href {\doibase
  10.1103/physrevlett.105.145302} {\bibfield  {journal} {\bibinfo  {journal}
  {Physical Review Letters}\ }\textbf {\bibinfo {volume} {105}},\ \bibinfo
  {pages} {145302} (\bibinfo {year} {2010})}\BibitemShut {NoStop}%
\bibitem [{\citenamefont {Souris}\ \emph {et~al.}(2011)\citenamefont {Souris},
  \citenamefont {Grucker}, \citenamefont {Dupont-Roc},\ and\ \citenamefont
  {Jacquier}}]{Souris2011}%
  \BibitemOpen
  \bibfield  {author} {\bibinfo {author} {\bibfnamefont {F.}~\bibnamefont
  {Souris}}, \bibinfo {author} {\bibfnamefont {J.}~\bibnamefont {Grucker}},
  \bibinfo {author} {\bibfnamefont {J.}~\bibnamefont {Dupont-Roc}}, \ and\
  \bibinfo {author} {\bibfnamefont {P.}~\bibnamefont {Jacquier}},\ }\href
  {\doibase 10.1007/s10909-010-0256-6} {\bibfield  {journal} {\bibinfo
  {journal} {Journal of Low Temperature Physics}\ }\textbf {\bibinfo {volume}
  {162}},\ \bibinfo {pages} {412} (\bibinfo {year} {2011})}\BibitemShut
  {NoStop}%
\bibitem [{End()}]{Endnote}%
  \BibitemOpen
  \href@noop {} {}\bibinfo {note} {Temperatures as low as 10 mK have been
  reached in unpublished work by the Balibar group, while 15 mK has been
  demonstrated in e.g. A. Haziot, X. Rojas, A.D. Fefferman, J.R. Beamish and S.
  Balibar, Physical Review Letters \textbf{110}, 035301 (2013).}\BibitemShut
  {Stop}%
\bibitem [{\citenamefont {Manninen}\ \emph {et~al.}(1992)\citenamefont
  {Manninen}, \citenamefont {Pekola}, \citenamefont {Kira}, \citenamefont
  {Ruutu}, \citenamefont {Babkin}, \citenamefont {Alles},\ and\ \citenamefont
  {Lounasmaa}}]{Manninen1992}%
  \BibitemOpen
  \bibfield  {author} {\bibinfo {author} {\bibfnamefont {A.~J.}\ \bibnamefont
  {Manninen}}, \bibinfo {author} {\bibfnamefont {J.~P.}\ \bibnamefont
  {Pekola}}, \bibinfo {author} {\bibfnamefont {G.~M.}\ \bibnamefont {Kira}},
  \bibinfo {author} {\bibfnamefont {J.~P.}\ \bibnamefont {Ruutu}}, \bibinfo
  {author} {\bibfnamefont {A.~V.}\ \bibnamefont {Babkin}}, \bibinfo {author}
  {\bibfnamefont {H.}~\bibnamefont {Alles}}, \ and\ \bibinfo {author}
  {\bibfnamefont {O.}~\bibnamefont {Lounasmaa}},\ }\href@noop {} {\bibfield
  {journal} {\bibinfo  {journal} {Physical Review Letters}\ }\textbf {\bibinfo
  {volume} {69}},\ \bibinfo {pages} {2392} (\bibinfo {year}
  {1992})}\BibitemShut {NoStop}%
\bibitem [{\citenamefont {Alles}\ \emph {et~al.}(1994)\citenamefont {Alles},
  \citenamefont {Ruutu}, \citenamefont {Babkin}, \citenamefont {Hakonen},
  \citenamefont {Manninen},\ and\ \citenamefont {Pekola}}]{Alles1994}%
  \BibitemOpen
  \bibfield  {author} {\bibinfo {author} {\bibfnamefont {H.}~\bibnamefont
  {Alles}}, \bibinfo {author} {\bibfnamefont {J.~P.}\ \bibnamefont {Ruutu}},
  \bibinfo {author} {\bibfnamefont {A.~V.}\ \bibnamefont {Babkin}}, \bibinfo
  {author} {\bibfnamefont {P.~J.}\ \bibnamefont {Hakonen}}, \bibinfo {author}
  {\bibfnamefont {A.~J.}\ \bibnamefont {Manninen}}, \ and\ \bibinfo {author}
  {\bibfnamefont {J.~P.}\ \bibnamefont {Pekola}},\ }\href {\doibase
  10.1063/1.1144827} {\bibfield  {journal} {\bibinfo  {journal} {Review of
  Scientific Instruments}\ }\textbf {\bibinfo {volume} {65}},\ \bibinfo {pages}
  {1784} (\bibinfo {year} {1994})}\BibitemShut {NoStop}%
\bibitem [{\citenamefont {Wagner}\ \emph {et~al.}(1994)\citenamefont {Wagner},
  \citenamefont {Ras}, \citenamefont {Remeijer}, \citenamefont {Steel},\ and\
  \citenamefont {Frossati}}]{Wagner1994}%
  \BibitemOpen
  \bibfield  {author} {\bibinfo {author} {\bibfnamefont {R.}~\bibnamefont
  {Wagner}}, \bibinfo {author} {\bibfnamefont {P.~J.}\ \bibnamefont {Ras}},
  \bibinfo {author} {\bibfnamefont {P.}~\bibnamefont {Remeijer}}, \bibinfo
  {author} {\bibfnamefont {S.~C.}\ \bibnamefont {Steel}}, \ and\ \bibinfo
  {author} {\bibfnamefont {G.}~\bibnamefont {Frossati}},\ }\href@noop {}
  {\bibfield  {journal} {\bibinfo  {journal} {Journal of Low Temperature
  Physics}\ }\textbf {\bibinfo {volume} {95}},\ \bibinfo {pages} {715}
  (\bibinfo {year} {1994})}\BibitemShut {NoStop}%
\bibitem [{\citenamefont {Pobell}(2007)}]{Pobell2007}%
  \BibitemOpen
  \bibfield  {author} {\bibinfo {author} {\bibfnamefont {F.}~\bibnamefont
  {Pobell}},\ }\href@noop {} {\emph {\bibinfo {title} {Matter and Methods at
  Low Temperatures}}},\ \bibinfo {edition} {3rd}\ ed.\ (\bibinfo  {publisher}
  {Springer-Verlag},\ \bibinfo {year} {2007})\BibitemShut {NoStop}%
\bibitem [{\citenamefont {Fujiwara}\ \emph {et~al.}(2012)\citenamefont
  {Fujiwara}, \citenamefont {Noda}, \citenamefont {Tanaka}, \citenamefont
  {Toubaru}, \citenamefont {Zhao},\ and\ \citenamefont
  {Takeuchi}}]{Fujiwara2012}%
  \BibitemOpen
  \bibfield  {author} {\bibinfo {author} {\bibfnamefont {M.}~\bibnamefont
  {Fujiwara}}, \bibinfo {author} {\bibfnamefont {T.}~\bibnamefont {Noda}},
  \bibinfo {author} {\bibfnamefont {A.}~\bibnamefont {Tanaka}}, \bibinfo
  {author} {\bibfnamefont {K.}~\bibnamefont {Toubaru}}, \bibinfo {author}
  {\bibfnamefont {H.-Q.}\ \bibnamefont {Zhao}}, \ and\ \bibinfo {author}
  {\bibfnamefont {S.}~\bibnamefont {Takeuchi}},\ }\href@noop {} {\bibfield
  {journal} {\bibinfo  {journal} {Optics Express}\ }\textbf {\bibinfo {volume}
  {20}},\ \bibinfo {pages} {19545} (\bibinfo {year} {2012})}\BibitemShut
  {NoStop}%
\bibitem [{\citenamefont {Arcizet}\ \emph {et~al.}(2009)\citenamefont
  {Arcizet}, \citenamefont {Rivi\`{e}re}, \citenamefont {Schliesser},
  \citenamefont {Anetsberger},\ and\ \citenamefont {Kippenberg}}]{Arcizet2009}%
  \BibitemOpen
  \bibfield  {author} {\bibinfo {author} {\bibfnamefont {O.}~\bibnamefont
  {Arcizet}}, \bibinfo {author} {\bibfnamefont {R.}~\bibnamefont
  {Rivi\`{e}re}}, \bibinfo {author} {\bibfnamefont {A.}~\bibnamefont
  {Schliesser}}, \bibinfo {author} {\bibfnamefont {G.}~\bibnamefont
  {Anetsberger}}, \ and\ \bibinfo {author} {\bibfnamefont {T.}~\bibnamefont
  {Kippenberg}},\ }\href {\doibase 10.1103/physreva.80.021803} {\bibfield
  {journal} {\bibinfo  {journal} {Physical Review A}\ }\textbf {\bibinfo
  {volume} {80}},\ \bibinfo {pages} {021803(R)} (\bibinfo {year}
  {2009})}\BibitemShut {NoStop}%
\bibitem [{\citenamefont {Hauer}\ \emph {et~al.}(2013)\citenamefont {Hauer},
  \citenamefont {Doolin}, \citenamefont {Beach},\ and\ \citenamefont
  {Davis}}]{Hauer2013}%
  \BibitemOpen
  \bibfield  {author} {\bibinfo {author} {\bibfnamefont {B.~D.}\ \bibnamefont
  {Hauer}}, \bibinfo {author} {\bibfnamefont {C.}~\bibnamefont {Doolin}},
  \bibinfo {author} {\bibfnamefont {K.~S.~D.}\ \bibnamefont {Beach}}, \ and\
  \bibinfo {author} {\bibfnamefont {J.~P.}\ \bibnamefont {Davis}},\ }\href
  {\doibase 10.1016/j.aop.2013.08.003} {\bibfield  {journal} {\bibinfo
  {journal} {Annals of Physics}\ }\textbf {\bibinfo {volume} {339}},\ \bibinfo
  {pages} {181} (\bibinfo {year} {2013})}\BibitemShut {NoStop}%
\end{thebibliography}%


\end{document}